\documentclass[apj,numberedappendix]{emulateapj}
\usepackage{amssymb,amsmath}
\usepackage{graphicx}
\usepackage{appendix}
\usepackage{natbib}
\usepackage{hyperref}
\usepackage{enumerate}
\usepackage{multirow}
\usepackage{bm}

\newcommand{\etal}{et~al.~}

\altaffiltext{\MIT}{Kavli Institute for Astrophysics and Space Research, Massachusetts Institute of Technology, 77 Massachusetts Avenue, Cambridge, MA 02139}
\altaffiltext{\Berkeley}{Department of Physics, University of California, Berkeley, CA 94720}
\altaffiltext{\FNAL}{Fermi National Accelerator Laboratory, Batavia, IL 60510-0500}
\altaffiltext{\KICPChicago}{Kavli Institute for Cosmological Physics, University of Chicago, 5640 South Ellis Avenue, Chicago, IL 60637}
\altaffiltext{\AAUChicago}{Department of Astronomy and Astrophysics, University of Chicago, 5640 South Ellis Avenue, Chicago, IL 60637}
\altaffiltext{\ANL}{Argonne National Laboratory, 9700 S. Cass Avenue, Argonne, IL, USA 60439}
\altaffiltext{\Miss}{Department of Physics and Astronomy, University of Missouri, 5110 Rockhill Road, Kansas City, MO 64110}
\altaffiltext{\Munich}{Faculty of Physics, Ludwig-Maximilians-Universit\"{a}t, Scheinerstr.\ 1, 81679 Munich, Germany}
\altaffiltext{\ExcellenceCluster}{Excellence Cluster Universe, Boltzmannstr.\ 2, 85748 Garching, Germany}
\altaffiltext{\MPE}{Max Planck Institute for Extraterrestrial Physics, Giessenbachstr.\ 85748 Garching, Germany}
\altaffiltext{\CfA}{Harvard-Smithsonian Center for Astrophysics, 60 Garden Street, Cambridge, MA 02138}
\altaffiltext{\UMontreal}{D\'{e}partement de Physique, Universit\'{e} de Montr\'{e}al, C.P. 6128, Succ. Centre-Ville, Montr\'{e}al, Qu\'{e}bec H3C 3J7, Canada}
\altaffiltext{\Huntingdon}{Huntingdon Institute for X-ray Astronomy, LLC}
\altaffiltext{\Melbourne}{School of Physics, University of Melbourne, Parkville, VIC 3010, Australia}
\altaffiltext{\IfA}{Institute for Astronomy (IFA), University of Hawaii, 2680 Woodlawn Drive, HI 96822}
\altaffiltext{\ill}{Department of Astronomy and Department of Physics, University of Illinois at Urbana-Champaign, 1002 W.\ Green Street, Urbana, IL 61801 USA}

\def\MIT{1}
\def\Berkeley{2}
\def\FNAL{3}
\def\KICPChicago{4}
\def\AAUChicago{5}
\def\ANL{6}
\def\Miss{7}
\def\Munich{8}
\def\ExcellenceCluster{9}
\def\CfA{10}
\def\UMontreal{11}
\def\Huntingdon{12}
\def\MPE{13}
\def\Melbourne{14}
\def\IfA{15}
\def\ill{16}

\begin{document}


\title{The Evolution of the Intracluster Medium Metallicity in Sunyaev-Zel'dovich-Selected Galaxy Clusters at $0 < \lowercase{z} < 1.5$}



\author{
M.~McDonald\altaffilmark{\MIT},
E.~Bulbul\altaffilmark{\MIT},
T.~de~Haan\altaffilmark{\Berkeley},
E.~D.~Miller\altaffilmark{\MIT},
B.~A.~Benson\altaffilmark{\FNAL,\KICPChicago,\AAUChicago},
L.~E.~Bleem\altaffilmark{\KICPChicago,\ANL},
M.~Brodwin\altaffilmark{\Miss},
J.~E.~Carlstrom\altaffilmark{\KICPChicago,\AAUChicago},
I.~Chiu\altaffilmark{\Munich,\ExcellenceCluster},
W.~R.~Forman\altaffilmark{\CfA},
J.~Hlavacek-Larrondo\altaffilmark{\UMontreal},
G.~P.~Garmire\altaffilmark{\Huntingdon},
N.~Gupta\altaffilmark{\Munich,\ExcellenceCluster,\MPE},
J.~J.~Mohr\altaffilmark{\Munich,\ExcellenceCluster,\MPE},
C.~L.~Reichardt\altaffilmark{\Melbourne},
A.~Saro\altaffilmark{\Munich,\ExcellenceCluster},
B.~Stalder\altaffilmark{\IfA},
A.~A.~Stark\altaffilmark{\CfA},
and J.~D.~Vieira\altaffilmark{\ill}
}

\email{Email: mcdonald@space.mit.edu}   


\begin{abstract}
We present the results of an X-ray spectral analysis of 153 galaxy clusters observed with the \emph{Chandra}, \emph{XMM-Newton}, and \emph{Suzaku} space telescopes. These clusters, which span $0 < z < 1.5$, were drawn from a larger, mass-selected sample of galaxy clusters discovered in the 2500 square degree South Pole Telescope Sunyaev Zel'dovich (SPT-SZ) survey.
With a total combined exposure time of 9.1 Ms, these data yield the strongest constraints to date on the evolution of the metal content of the intracluster medium (ICM).
We find no evidence for strong evolution in the global ($r<R_{500}$) ICM metallicity ($dZ/dz = -0.06 \pm 0.04 Z_{\odot}$), with a mean value at $z=0.6$ of $\left<Z\right> = 0.23 \pm 0.01$ Z$_{\odot}$ and a scatter of $\sigma_Z = 0.08 \pm 0.01$ Z$_{\odot}$. These results imply that the emission-weighted metallicity has not changed by more than 40\% since $z=1$ (at 95\% confidence), consistent with the picture of an early ($z>1$) enrichment. 
We find, in agreement with previous works, a significantly higher mean value for the metallicity in the centers of cool core clusters versus non-cool core clusters. 
We find weak evidence for evolution in the central metallicity of cool core clusters ($dZ/dz = -0.21 \pm 0.11 Z_{\odot}$), which is sufficient to account for this enhanced central metallicity over the past $\sim$10 Gyr.
We find no evidence for metallicity evolution outside of the core ($dZ/dz = -0.03 \pm 0.06 Z_{\odot}$), and no significant difference in the core-excised metallicity between cool core and non-cool core clusters. This suggests that strong radio-mode AGN feedback does not significantly alter the distribution of metals at $r>0.15R_{500}$.
Given the limitations of current-generation X-ray telescopes in constraining the ICM metallicity at $z>1$, significant improvements on this work will likely require next-generation X-ray missions.
\end{abstract}

\keywords{galaxies: clusters: general -- galaxies: clusters: intracluster medium -- X-rays: galaxies: clusters \vspace{-0.2in}}

\section{Introduction}
\setcounter{footnote}{0}

Galaxy clusters, which are the most massive collapsed structures in the Universe, are made up of hundreds to thousands of galaxies, a massive reservoir of hot ($\gtrsim$10$^7$K) plasma, and dark matter. The latter dominates the mass budget, contributing $\sim$90\% of the the total mass for a massive galaxy cluster \citep[e.g.,][]{chiu14}. Of the remaining $\sim$10\% in mass, the hot intracluster medium (ICM) outweighs the stars by a factor of roughly ten \citep[e.g.,][]{lin03b,gonzalez13}. Visible in X-rays as diffuse emission on Mpc scales, the hot ICM retains the imprint of the cluster history in its thermodynamic and chemical abundance profiles and in its X-ray morphology, allowing major events in the history of a given cluster to be inferred billions of years later.

The chemical abundance of the hot ICM contains the cumulative enrichment history from various processes since the Big Bang, including mass loss from evolved stars, heavy element enrichment from supernovae, and dilution of metals due to mixing with low-metallicity gas infalling along cosmic filaments. In practice, with CCD-resolution X-ray spectra from \emph{Chandra} and \emph{XMM-Newton}, constraints on the ICM metal abundance come primarily from the equivalent width measurement of the Fe K$\alpha$ emission line at 6.7 keV. Assuming that the ratio of iron to other elements is constant across the Universe, we can infer a metal abundance by comparing the iron abundance in a given cluster to that of the Sun. In massive, low-$z$ clusters ($z < 0.3$), the average observed metallicity is roughly a third of the solar value \citep[e.g.,][]{serlemitsos77,arnaud92,mushotzky96, mushotzky97,degrandi01,degrandi04,baldi07,leccardi08b,sanderson09,matsushita11, bulbul12b,molendi15}, assuming solar abundances from \cite{anders89}. In ``cool core'' clusters -- those with central cooling times significantly shorter than the age of the Universe \citep[e.g.,][]{hudson10} -- the metallicity peaks in the center \citep[e.g.,][]{degrandi01,degrandi04,baldi07,leccardi08b,johnson11,elkholy15}, sometimes reaching values $Z > Z_{\odot}$ \citep[e.g.,][]{kirkpatrick15}. This may be owing to the fact that the central, most massive galaxy (which can enrich the ICM via stellar mass loss) tends to be in the cluster center for cool core clusters, leading to centrally-peaked stellar-to-gas ratios for these systems which is not present in non-cool core clusters.

The metal enrichment history of the ICM remains a mystery.  Recent studies \citep[e.g.,][]{balestra07,maughan08, anderson09,andreon12,baldi12b, ettori15} have attempted to quantify how the metallicity evolves, both in the core and outer regions, for galaxy clusters at $0 \lesssim z \lesssim 1$. Using data from the \emph{Chandra X-ray Observatory}, \cite{maughan08} reported evolution of the metal abundance for 115 galaxy clusters at $z < 1.3$. The sample, defined as all known clusters with existing data in the \emph{Chandra} archive at the time, is fairly representative of the true cluster population at low redshift, but is biased towards extreme (well-studied) systems at high redshift. Further, it is known that some of the most relaxed systems (which exhibit enhanced metallicity in their cores) have been missed by X-ray surveys because their cool cores appear point-like at large distances \citep[e.g., the Phoenix cluster;][]{mcdonald12c}. The sample used by \cite{maughan08} had few (12) clusters at $z>0.7$, where they observed the strongest evolution. More recent studies, first by \cite{baldi12b} and then \cite{ettori15}, reported results from the \emph{XMM-Newton} telescope, employing a sample of all clusters in the archive at $z>0.4$. This sample likely suffers from similar selection biases to the study by \cite{maughan08}. These later works find no measurable evolution in the global metallicity of clusters at $z<1.39$, although \cite{ettori15} find marginal ($>2\sigma$) evidence for evolution in the centers of cool core clusters. None of these studies find a strong dependence between cluster mass (or temperature) and metallicity, with \cite{baldi12b} reporting $Z \propto kT^{0.06 \pm 0.16}$. 

Here, we present the first study of ICM metallicity evolution in a sample of galaxy clusters selected via the Sunyaev Zel'dovich effect \citep{sunyaev72}. This selection yields a mass-limited, redshift-independent sample of clusters that is relatively unbiased to the presence or lack of a cool core \citep{lin15}. We focus here on X-ray follow-up of galaxy clusters at $0 < z < 1.2$, providing the best constraints on the global metallicity evolution and the radius-dependent evolution over this redshift range. The data used in this analysis, from the \emph{Chandra}, \emph{XMM-Newton}, and \emph{Suzaku} telescopes, are described in \S2. In \S3 we present the constraints on metallicity evolution, and describe the dependence of this evolution on the presence or lack of a cool core. In \S4 we speculate on the cause of metal enrichment, both in the cores and outskirts of galaxy clusters, and compare our results to previous works. Finally, in \S5 we will summarize the results and comment on the ability of future surveys to improve these constraints.

Throughout this work we assume H$_0$ = 70 km s$^{-1}$ Mpc$^{-1}$, $\Omega_M$ = 0.3, $\Omega_{\Lambda}$ = 0.7.

\section{Data \& Analysis}
\subsection{The South Pole Telescope Cluster Survey}

\begin{figure}[htb]
\centering
\includegraphics[width=0.49\textwidth]{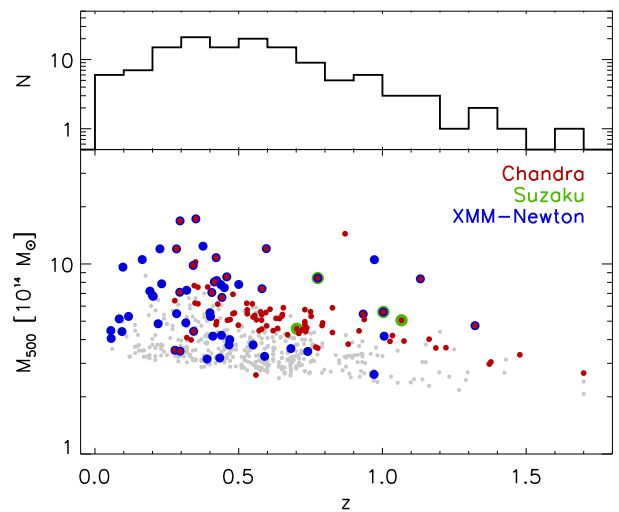}
\caption{Distribution of cluster masses (M$_{500}$) versus redshift for 516 clusters discovered in the full SPT-SZ survey \citep[grey points;][]{bleem15}. Clusters observed with \emph{Chandra}, \emph{XMM-Newton}, and \emph{Suzaku} are colored red, blue, and green, respectively. This figure demonstrates the clean mass selection of this sample. Masses in this plot are estimated from the SZ significance, following \cite{bleem15}.}
\label{fig:Mz}
\end{figure}

This work is based on a sample of galaxy clusters selected via the Sunyaev Zel'dovich (SZ) effect in the 2500 square degree South Pole Telescope Sunyaev Zel'dovich (SPT-SZ) Survey \citep{bleem15}. This survey, completed in 2011, discovered 516 galaxy clusters in the southern sky, at $0 < z \lesssim 1.7$ with masses $M_{500} \gtrsim 3 \times 10^{14}$ M$_{\odot}$. For a more detailed description of the selection and cluster properties, the reader is directed to \cite{bleem15}. For a subsample of 153 clusters we have X-ray follow-up from the \emph{Chandra}, \emph{XMM-Newton}, and/or \emph{Suzaku} telescopes. The mass-redshift distribution for the clusters with follow-up X-ray data is shown in Figure \ref{fig:Mz}. This figure emphasizes the uniform mass selection and broad redshift coverage of our sample. Below, we describe in detail the follow-up strategy and data analysis for each of these three X-ray observatories.

\subsection{SPT--Chandra Sample}
From the larger sample of 516 confirmed clusters in the SPT-SZ survey, we have performed a nearly mass-limited X-ray follow-up of 96 clusters at $0.25 < z < 1.5$ and M$_{500} \gtrsim 4 \times 10^{14}$ M$_{\odot}$ with the \emph{Chandra X-ray Observatory}. The majority of these clusters were observed as part of either an \emph{X-ray Visionary Program} to obtain shallow X-ray imaging of the 80 most massive SPT-selected clusters at $z>0.3$ (PI: Benson) or a Large Program to observe 10 SPT-selected clusters at $z>1.2$ (PI: McDonald). The remaining systems were observed through various smaller GO (PIs: McDonald, Mohr) and GTO (PIs: Garmire, Murray) programs, or were available in the archive. The details of these observations are summarized in \cite{mcdonald13b,mcdonald14c}.

For each SPT-selected cluster observed with \emph{Chandra}, the total exposure time was estimated with a goal of obtaining 2000 X-ray counts, assuming relationships between the SZ detection significance and cluster mass, and between cluster mass and X-ray luminosity \citep{andersson11}. Owing to significant scatter in these relations, coupled with uncertainty in preliminary cluster redshifts, the observed X-ray counts vary by a factor of $\sim$2.5 around this goal \citep[see Fig.\ 2 in][]{mcdonald14c}. 
We have much deeper data ($\gg$10,000 counts) on a subset of 7 well-studied clusters in this sample (e.g., Bullet Cluster (PI: Markevitch), El Gordo (PI: Hughes), Phoenix (PI: McDonald)), which yields well-measured metallicities in a subsample of systems. We note that, since we consider both the intrinsic scatter and measurement uncertainty in our fitting procedure, the inclusion of high-S/N systems will not strongly bias our results. We will return to this point in \S5.2.

For each cluster we estimate R$_{500}$, the radius within which the average enclosed density is 500 times the critical density ($\rho_{crit}$), based on the SZ-derived mass presented in \cite{bleem15}. We note that SZ-derived masses may be underestimated for low-$z$ ($z < 0.25$) clusters and are only approximate for low-significance ($\xi < 5$) clusters \citep{bleem15}. However, this already-small bias has a reduced ($M^{1/3}$) effect on the inferred radius, which has an even smaller effect on the integrated metallicity, given the flatness of the metallicity profile at $r \sim R_{500}$ \citep{leccardi08b}.

\subsubsection{Chandra Data Reduction}
X-ray data from \emph{Chandra} are reduced following procedures outlined in \cite{vikhlinin05}, \cite{andersson11}, and \cite{mcdonald13b,mcdonald14c} using \textsc{ciao} v4.7 and \textsc{caldb} v4.6.8, along with the latest blank-sky background and response files. All exposures are initially filtered for background flares, before applying the latest calibration corrections and determining the appropriate (epoch-based) blank-sky background. Given that the typical angular size of a galaxy cluster at $z=0.6$ is only $\sim$2.5$^{\prime}$, we are also able to extract off-source background files at distances of $>$3R$_{500}$ from the cluster center for each observation. Blank-sky background spectra are rescaled based on the observed 9.5--12\,keV flux, and subtracted from both on- and off-source spectra. Any residual background (e.g., from unresolved point sources or galactic emission) is accounted for by simultaneous modeling of on- and off-source spectra (see below).
Point sources are identified using an automated routine following a wavelet decomposition technique \citep{vikhlinin98}, and then visually inspected. The center of the cluster is chosen by iteratively measuring the centroid in a 250--500~kpc annulus, following \cite{mcdonald13b}. This choice of center is insensitive to structure in the core and is a more accurate proxy for the center of the underlying dark matter potential than the X-ray peak.

\subsubsection{Chandra Spectral Fitting}
For each cluster observation, we extract spectra in three different annuli: i) $0 < r < R_{500}$ (total); ii) $0 < r < 0.15R_{500}$ (core); and iii) $0.15R_{500} < r < R_{500}$ (core-excised). The first of these annuli will be compared with measurements from \emph{XMM-Newton} and \emph{Suzaku}, which, for the high-$z$ systems ($z>1$), do not have sufficient angular resolution to excise the core or identify whether the cluster harbors a cool core or not. The combination of full-aperture data from all three telescopes will allow us to tightly constrain the evolution of the total metal content in clusters out to $z\sim1$. The latter two annuli, which we can only measure using the high angular resolution \emph{Chandra} data, will provide weaker constraints on how this evolution depends on the presence or lack of a cool core, and whether it is stronger in the core than in the outskirts.

\begin{figure}[t]
\centering
\includegraphics[width=0.49\textwidth]{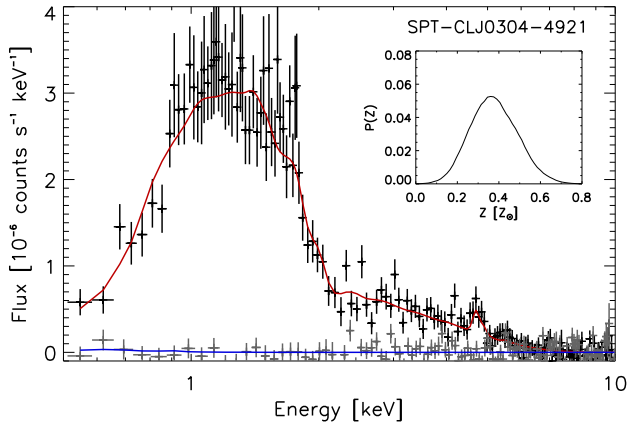}
\caption{\emph{Chandra} X-ray spectrum of SPT-CLJ0304-4921. Black points show the on-source spectrum, integrated within R$_{500}$, and grey points show the off-source background, normalized to the same area. Red and blue curves show the best-fit models for the on- and off-source spectra, respectively. The presence of a redshifted iron emission line at $\sim$5 keV allows us to constrain the metallicity of this system -- these constraints from 10,000 MCMC chains are shown in the inset (see \S2.2.2 for more details).}
\label{fig:spectrum}
\end{figure}

Spectra are individually fit in combination with their respective background spectra (see \S2.2.1) over the energy range 0.5--10.0~keV with \textsc{xspec} \citep[v12.9.0;][]{arnaud96}, using a combination of a single-temperature plasma \citep[\textsc{apec};][]{smith01}, a soft X-ray Galactic background (\textsc{apec}, $kT=0.18$~keV, $Z = Z_{\odot}$, $z=0$), a hard X-ray cosmic background (\textsc{bremss}, $kT=40$~keV), and a Galactic absorption model (\textsc{phabs})\footnote{\scriptsize http://heasarc.gsfc.nasa.gov/docs/xanadu/xspec/manual/ \\ XspecModels.html}. Metallicity measurements are based on the solar abundances of \cite{anders89}. For each observation, the normalization of the hard X-ray background is tied between the on- and off-source spectra, while the soft X-ray background is tied between all spectra for a given cluster (i.e., multiple OBSIDs). This difference is to account for the fact that the hard X-ray background should be exposure time dependent (i.e., larger fraction of the cosmic X-ray background (CXB) resolved into point sources for long exposure times), while the soft, diffuse X-ray background is not. An example result of this fitting procedure is shown in Figure \ref{fig:spectrum}.

The inferred ICM metallicity is most sensitive to the Fe~K$\alpha$ emission line at 6.7 keV. Since the fit to this line is also strongly dependent on the cluster redshift, we allow the redshift to float in the fitting procedure within the 2$\sigma$ uncertainties quoted by \cite{bleem15}. For photometric redshifts, this uncertainty is typically $\Delta z \sim 0.02(1+z)$. For spectroscopic redshifts, we use a fixed uncertainty of 3000 km s$^{-1}$ ($\Delta z \sim 0.01$). In a similar fashion, we allow the Galactic absorption column to float within $\sim$15\% of the measured value from \cite{kalberla05}, which tends to improve the continuum fit. Goodness-of-fit is determined using a modified version of the $\chi^2$ parameter, following \cite{churazov96}, which has been shown to yield unbiased parameter estimates for spectra containing as few as $\sim$50 total counts. 

\begin{figure}
\centering
\includegraphics[width=0.49\textwidth]{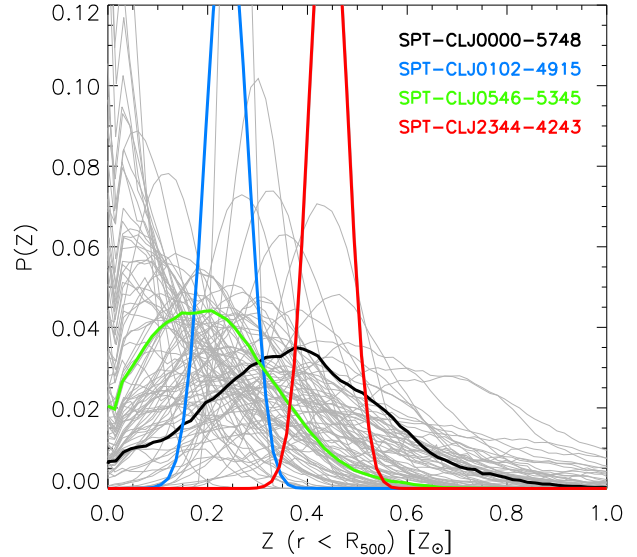}
\caption{Probability distribution functions for the ICM metallicity ($P(z)$) at $r<R_{500}$ for 96 SPT-selected clusters that have been observed with \emph{Chandra}. These curves were derived using the Monte Carlo Markov Chain solver in \textsc{xspec}, as described in \S2.2.2. We highlight four clusters in color, which demonstrate weak (green, black) and strong (blue, red) constraints for low-metallicity (blue, green) and high-metallicity (black, red) clusters.}
\label{fig:fig1}
\end{figure}

For each cluster, we use the Markov-Chain Monte-Carlo (MCMC) solver in \textsc{xspec} to determine the full probability distribution for the ICM metallicity. Using the Metropolis-Hastings algorithm, we run 5 chains for each cluster, each with a length of 25000 and discarding the first 5000 steps. Integrating over all other free parameters ($n_H$, $kT$, $z$, source normalization, soft background normalization, hard background normalization) yields the posterior distribution for $Z$ (hereafter $P(Z)$). In Figure \ref{fig:fig1}, we show the derived $P(Z)$ for each cluster, highlighting two non-cool core clusters (hereafter NCC; SPT-CLJ0102-4915, SPT-CLJ0546-5345) and two cool-core clusters (hereafter CC; SPT-CLJ0000-5748, SPT-CLJ2344-4243), where we use the central entropy ($K \equiv kTn_e^{-2/3}$) as the classification metric, following \cite{mcdonald13b}. For both types (CC/NCC), we show a low-S/N case and a high-S/N case, to demonstrate the difference in metallicity constraints from system to system. In \S3, we will describe how these individual cluster constraints are incorporated into a framework which allows us to constrain the overall metallicity evolution.

%

\subsection{XMM-Newton Data and Spectral Fitting}

A significant number of SPT-selected clusters have been observed by \emph{XMM-Newton} as part of several different programs including, but not limited to, those focused on weak lensing mass calibration (PIs: Benson, Suhada) and the study of the highest redshift SPT-selected systems (PIs: Benson, Andersson). In all, as of the end of 2015, 69 SPT-selected clusters have been observed by \emph{XMM-Newton}, with no preference in selection for the dynamical/cooling state of the cluster -- the full details of this sample are provided in Bulbul \etal (in prep). Given that the fraction of cool cores in this randomly-selected subsample should be representative of the true population, we are justified in combining these data with the larger, more complete sample of 96 clusters observed with \emph{Chandra}.

Event files were calibrated using the XMM-Newton Science Analysis System (SAS) version 14.0.0, and the most recent calibration files as of October 2015. Calibrated, clean event files were produced after filtering for high intensity particle-background flares. Additional details of data reduction and analysis are described in detail in \cite{bulbul12}.

The images are extracted in a 0.4--7.0 keV bandpass from all instruments and pointings. The particle background and soft proton background subtracted images were used to detect the point source in the field of view. Detected point sources were excluded from the further analysis. The source and background spectra were extracted within the overdensity radius R$_{500}$, as derived from \cite{bleem15}. We model the background with a superposition of four main components: quiescent particle background, cosmic X-ray background emission (including Galactic halo, local hot bubble, and unresolved extragalactic sources), solar wind charge exchange, and residual contamination from soft protons. We use the \emph{ROSAT} All-Sky Survey background spectrum which  is extracted from an annulus from 0.5$^\circ$ to 1.5$^\circ$  surrounding the cluster center to model the soft X-ray background. The 0.3--10 keV energy interval is used for MOS spectra, whereas the 0.4 -10.0 keV band was used for the PN fits. The remaining cluster model parameters and fitting process was identical to that used in the \emph{Chandra} analysis.

\subsection{Suzaku Data and Spectral Fitting}

Four SPT-selected clusters were observed with \emph{Suzaku} during Cycle 6 (PI: E.\ Miller). One of the primary goals of this program was to constrain the metallicity of galaxy clusters at high redshift, with the four clusters spanning $0.7 < z < 1.1$ (see Figure \ref{fig:Mz}). These systems were found to have slightly higher than average metallicities, with values ranging from 0.26--0.45 Z$_{\odot}$ \citep{miller12}. However, with only four clusters, we were unable to make any claim on metallicity evolution with certainty. Thus, we have included these four high-S/N measurements into a larger sample, which provide marginal improvements on the overall metallicity evolution of SPT-selected clusters.

Event data for XIS0, XIS1, and XIS3 were reprocessed using the most current \emph{Suzaku} calibration products as of September 2015.  Point sources were identified from shallow \emph{Chandra} data for each cluster using the CIAO tool \textsc{wavdetect}, and masked out in the \emph{Suzaku} data to a radius of 1$^{\prime}$.  Spectra for cluster emission were extracted from a region 4$^{\prime}$ in radius centered on the X-ray peak (about twice the typical R$_{500}$ for these four clusters) to account for the large size (2$^{\prime}$ half-power diameter) of the \emph{Suzaku} PSF.  This is sufficient to collect 99\% of the counts from an on-axis point source.  X-ray background spectra were extracted from the remaining source-free regions of each detector, excluding the central 5$^{\prime}$, the calibration source regions, and bad detector areas.

Response files were produced for the source and X-ray background with \textsc{xisrmfgen} and \textsc{xissimarfgen}, the latter using a background-subtracted \emph{Chandra} image of the cluster as an input source for ray-tracing to produce the cluster auxiliary response file (ARF).  The X-ray background ARF was constructed using a uniform, 20$^{\prime}$ radius source.  Particle background spectra were produced for each source and X-ray background region with the FTOOL \textsc{xisnxbgen}, with a filter to exclude parts of the spacecraft orbit with low geomagnetic cut-off rigidity (COR2 $>$ 6 GV) and thus reduce the background.  This filter was also applied to the source and X-ray background spectra.

Spectral fitting was performed with XSPEC v12.9, fitting all three XIS simultaneously with all free paramaters tied between instruments.  The particle background spectra were subtracted from the source and X-ray background spectra during spectral fitting.  The X-ray background model consisted of a one solar-abundance APEC component to model the total Galactic foreground, with temperature and normalization free; and one power law component with fixed $\Gamma = 1.4$ and free normalization to model the unresolved cosmic X-ray background.  Both components were absorbed by the same $N_H$ column as the cluster model.  The X-ray background parameters were tied between the cluster and background regions.  The remaining cluster model parameters and fitting process was identical to that used in the \emph{Chandra} analysis.


\section{Non-Parametric Metallicity Evolution}

\begin{figure}[b]
\centering
\includegraphics[width=0.49\textwidth]{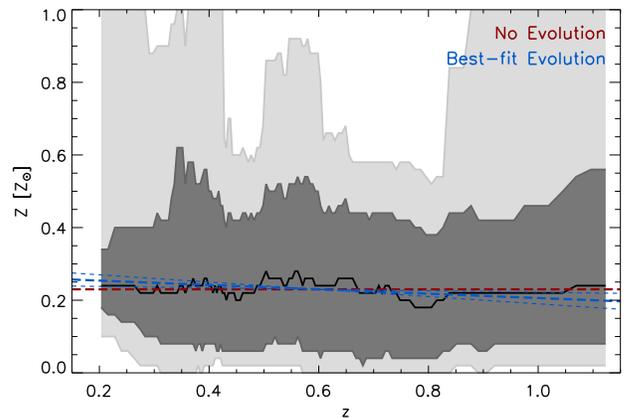}
\caption{Median metallicity (within $R_{500}$) as a function of redshift. For each redshift bin, we combine the probability functions, $P(z)$, for the nearest 21 clusters in redshift space, computing the combined probability distribution for the metallicity of clusters at that redshift. In black, we show the median value of that distribution, which does not appear to evolve with redshift. In dark and light grey we show the $1\sigma$ (68\%) and $2\sigma$ (95\%) confidence regions, respectively. The horizontal red line shows a non-evolving metallicity, while the blue lines show the best-fit model from the following section. This figure demonstrates that both the scatter and median values are well-fit by a non-evolving model, justifying our use of a simple, linear evolution term in \S4.}
\label{fig:Zz_allxray}
\end{figure}

The ultimate goal of this work is to quantify how the integrated ICM metallicity has evolved in galaxy clusters over the past $\sim$9 Gyr. As a first step, we show in Figure \ref{fig:Zz_allxray} the evolution in the ICM metallicity for subsamples of 21 clusters from $z\sim0.2$ to $z\sim1.2$, based on the full sample of clusters from all three X-ray observatories. In each redshift bin, we combine the P(Z) distributions for the 21 nearest clusters in redshift space, showing the median value of this combined probability distribution, as well as the 1$\sigma$ (dark grey) and 2$\sigma$ (light grey) widths. This figure provides three initial insights into the evolution of the global ($0 < r <$~R$_{500}$) metal content of the ICM. First, there is no sign of complex (e.g., non-linear) evolution in the median metallicity. Given the quality of these data, any evolution that does exist would be as well fit by a linear term in redshift (e.g., $Z \propto z$) as it would by something more complex (e.g., $Z \propto e^{-t/\tau}$). Second, the scatter in metallicity at a given redshift appears to be roughly symmetric about the median, such that it is well-approximated by a Gaussian distribution with $\sigma \sim 0.15$Z$_{\odot}$. There does appear to be some asymmetry to the distribution (because $Z<0$ is unphysical), but this asymmetry is not present at the 1$\sigma$ level. Finally, there does not appear to be any significant evolution in the scatter about the median metallicity out to $z\sim1$.

The combination of these three insights implies that a simple linear evolution (in redshift) with non-evolving gaussian scatter is an appropriate choice of model given our data quality and the properties of the clusters in this sample. In the following section, we will employ this simple model to provide constraints on the evolution of the average ICM metallicity as a function of both redshift and radius.


\section{Maximum Likelihood Analysis of Metallicity Evolution}

Given the large uncertainties inherent in our metallicity measurements, and the justification provided in the previous section, we opt for a simple linear model to express the evolution of metallicity as a function of redshift:

\begin{equation}
Z(z^{\prime}) = a + bz^{\prime} ~,
\end{equation}

\noindent{}where $z^{\prime} = z - 0.6$ is chosen to roughly minimize covariance between $a$ and $b$ for the subsample of clusters observed with \emph{Chandra} ($\left<z\right> \sim 0.6$). We note that, while the data presented in this paper are of sufficient quality to constrain the two parameters in this simple model, they are not of sufficient quality to determine if a more complex functional form is justified, especially at $z>1$. Assuming Gaussian intrinsic scatter (independent of metallicity and redshift), characterized by a width c, the probability that a cluster at redshift $z$ has metallicity $Z$ is given by:

\begin{equation}
P^{\prime}(Z|z,a,b,c) = \frac{1}{\sqrt{2\pi c^2}}\exp\left(-\frac{(Z - (a+bz^{\prime})^2}{2c^2}\right)~.
\end{equation}

\noindent{}Combining these probabilities for the full sample yields the posterior probability for the set of model parameters:

\begin{equation}
P(a,b,c|z,Z) \propto P(a,b,c)\prod\limits_{i} \int{dZ_iP^{\prime}(Z_i | z_i,a,b,c)P(Z_i)}~,
\end{equation}

\noindent{}Where $i$ refers to a given cluster, $P(a,b,c)$ is the prior probability (taken to be flat for all parameters), and $P(Z_i)$ is the probability distribution for the metallicity of a given cluster, as obtained from \textsc{xspec} (e.g., Figure \ref{fig:fig1}). We can then write the approximate likelihood as:

\begin{equation}
P(a,b,c|z,Z) \propto \prod\limits_{i} \sum\limits_{j}^{N} P^{\prime}(Z_{i,j} | z_i,a,b,c)~,
\end{equation}

\noindent{}where $Z_{i,j}$ denotes the $j^{th}$ sample (from $N=10^5$ total samples) drawn from the posterior $Z$ distribution of the $i^{th}$ cluster, as given by the MCMC solver in \textsc{xspec}. This posterior probability distribution of the model parameters is then explored using a uniformly spaced fine grid in all three parameters. We note that, by including intrinsic scatter in this fit, we reduce the effect that a few high S/N systems can have on the fit outcome. This approach is similar to considering a modified $\chi^2$ of the form $\Sigma (x_i - \mu)^2 / (\sigma_i^2 + \sigma_{intrinsic}^2)$.

\begin{figure}[htb]
\centering
\includegraphics[width=0.49\textwidth]{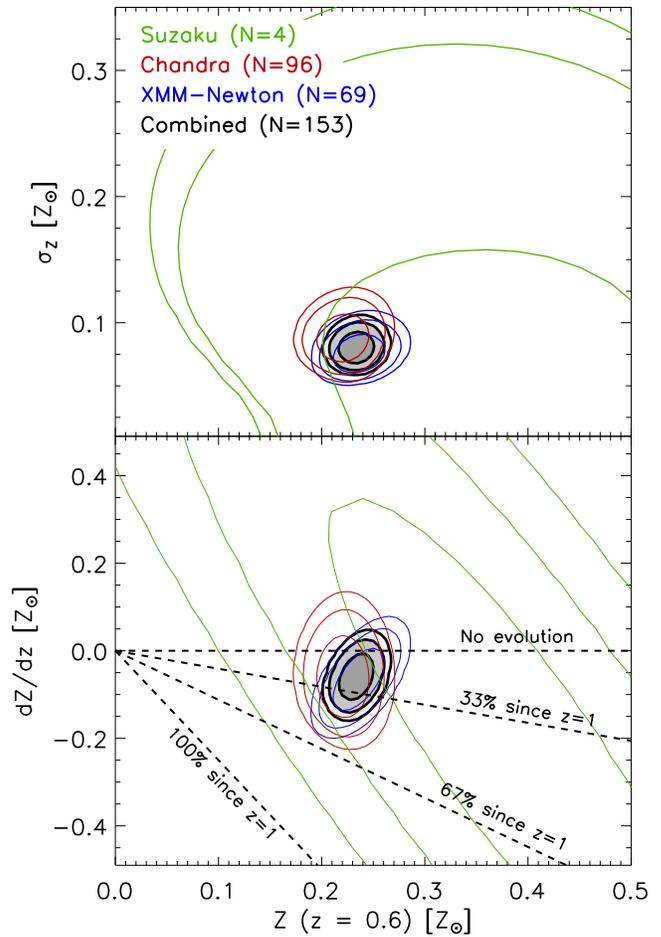}
\caption{Constraints on the evolution of the total metallicity within $R_{500}$. In the upper panel, we show the constraints on the scatter and normalization, while in the lower panel we compare the slope and normalization terms. In both panels, we show constraints from each of the three telescopes individually, along with the combined constraints in grey. Contours represent confidence levels of 68\%, 95\%, and 99.7\%. Overplotted in black dashed lines is a grid of evolutionary scenarios, to help contextualize these results. The tight constraints on the evolutionary term imply that, at most (with $>$95\% confidence), 40\% of the intracluster metals (including the core) in present-day clusters were created in the past $\sim$8 Gyr. This suggests that the bulk of the metals in the ICM were most likely created very rapidly early on.}
\label{fig:mcmc_allxray}
\end{figure}

Applying this formalism to the data described in \S2 yields posterior distributions for $a,b,c$, which we now assign the more physically-appropriate labels of $\left<Z[z=0.6]\right>$, $dZ/dz$, and $\sigma_Z$, respectively. In Figure \ref{fig:mcmc_allxray}, we show the constraints on each of these three parameters for the full set of 153 clusters with \emph{Chandra}, \emph{XMM-Newton}, and/or \emph{Suzaku} data. Here we are considering the evolution of the \emph{global} metallicity ($r<R_{500}$). For this sample, we find $\left<Z[z=0.6]\right> = 0.23 \pm 0.01Z_{\odot}$, $dZ/dz = -0.06 \pm 0.04 Z_{\odot}$, and $\sigma_Z = 0.08 \pm 0.01 Z_{\odot}$ -- this best-fitting model is overplotted on the data in Figure \ref{fig:Zz_allxray}. The 1$\sigma$ contours for all three parameters overlap between subsamples, suggesting that there is no systematic difference in either selection or treatment of the data.

Figure \ref{fig:mcmc_allxray} indicates that there is very little (if any) evolution in the global ICM metallicity over the past $\sim$9 Gyr.  These data indicate that, in a typical galaxy cluster with $M_{500} \gtrsim 3\times10^{14}$ M$_{\odot}$, the emission-weighted metallicity has changed by less than 40\% since $z=1$ ($>$95\% confidence). Such a lack of evolution is consistent with previous studies, with this work providing the strongest constraints to date. One can ask whether the evolution is stronger in the cores of clusters, where baryonic processes are important, than in the cluster outskirts, or whether the dynamical state of the cluster (e.g., relaxed or unrelaxed) has any bearing on the observed metallicity. To address both of these questions, we require the high angular resolution provided by \emph{Chandra} and, thus, will limit the remaining analysis in this section to the 96 clusters with \emph{Chandra} observations.

\begin{figure*}
\centering
\includegraphics[width=0.99\textwidth]{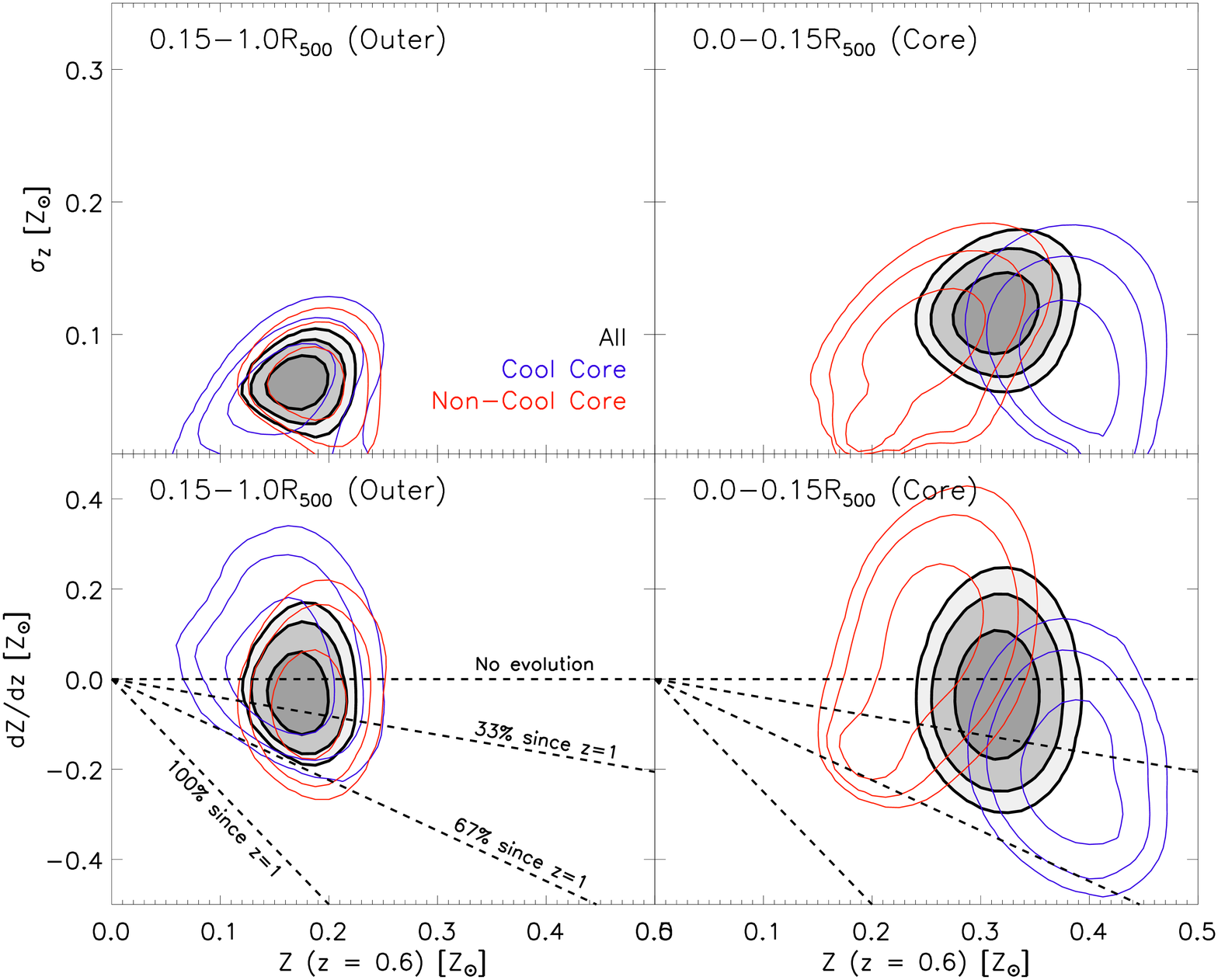}
\caption{Constraints on the evolution of the core ($r < 0.15R_{500}$) and outer ($r>0.15R_{500}$) metallicity. Here we have restricted our analysis to include only \emph{Chandra} data, where the core can be properly isolated for high-$z$ systems. We show individually constraints from cool core ($K_0 < 100$ keV cm$^2$) and non-cool core ($K_0 > 100$ keV cm$^2$) clusters. Outside of the core, clusters are remarkably similar, with relatively small scatter ($0.06 \pm 0.01$ Z$_{\odot}$) and no evidence for evolution. The metallicity outside of the core appears to be independent of the thermodynamic state of the core, suggesting that strong AGN outbursts, which are found predominantly in cool core clusters, are not significantly influencing the large-scale metallicity. We find significant dependence of the core metallicity on the presence or lack of a cool core, with cool core clusters having significantly higher normalization, while also showing marginal evidence (2$\sigma$) of evolution.}
\label{fig:mcmc_chandra}
\end{figure*}

The measured evolution of the \emph{spatially-resolved} ICM metallicity for clusters observed with \emph{Chandra} is shown in Figure \ref{fig:mcmc_chandra} and summarized in Table \ref{table:mcmc}. As described in \S2, we consider the metallicity in the core volume ($r<0.15$R$_{500}$), the core-excised volume (0.15R$_{500} < r < $ R$_{500}$), and the full volume ($r <$ R$_{500}$). Figure \ref{fig:mcmc_chandra} shows that these data, despite being of relatively low quality on a per-cluster basis ($\sim$2000 X-ray counts per cluster), provide excellent constraints on the metallicity evolution. We exclude \emph{XMM-Newton} and \emph{Suzaku} data from this part of the analysis, due to the fact that, at $z>0.5$, $>$20\% of the flux from the core will be scattered outside of 0.15R$_{500}$. Given the broad redshift range covered, disentangling this PSF-driven bias in the core-excised metallicity from an actual evolution would be challenging.

We find that the core-excised metallicity has a mean value of $\left<Z[z=0.6]\right> = 0.17 \pm 0.02$ Z$_{\odot}$ and $dZ/dz = -0.03 \pm 0.06$, corresponding to a mean value at redshift zero of $\left<Z[z=0.0]\right> = 0.19 \pm 0.04$ Z$_{\odot}$. This is remarkably consistent with the measured value from deep observations of the Perseus cluster of $0.212 \pm 0.008$ Z$_{\odot}$ \citep{werner13}, and other nearby clusters \citep{leccardi08b}, assuming solar abundances from \cite{anders89}. We find no evidence for evolution in the metallicity outside of the core, independent of whether the cluster is a cool core or non-cool core, suggesting that the bulk of the metals outside of the core were created at $z>1$. Again, this is consistent with earlier work by \cite{werner13}, who showed that the lack of azimuthal variations in the metallicity of the Perseus cluster implies that metal enrichment happened very early in the cluster lifetime. We note that there is very little intrinsic scatter in the core-excised metallicity ($\sigma_Z = 0.06 \pm 0.01$ Z$_{\odot}$), and that the mean value, redshift slope, and scatter of the metallicity outside of the core is independent of the cooling state (entropy) of the core.

In the cores of galaxy clusters ($r < 0.15$R$_{500}$), we find, for the full population, significantly higher metallicity. The mean value of $\left<Z[z=0.6]\right> = 0.32 \pm 0.03$ Z$_{\odot}$ is offset from the core-excised value by 5$\sigma$. Further, if we divide the sample roughly in half into CC (K$_0 < 100$ keV cm$^2$) and NCC (K$_0 > 100$ keV cm$^2$) subsamples, the mean metallicity for each subsample is $\left<Z[z=0.6]\right>_{CC} = 0.39 \pm 0.03$ Z$_{\odot}$ and $\left<Z[z=0.6]\right>_{NCC} = 0.24 \pm 0.05$ Z$_{\odot}$, respectively. That is, there is a $\sim$4$\sigma$ offset between the core metallicity in CC and NCC clusters. Despite this offset, we measure no difference in the scatter of core metallicity between CC and NCC systems. This dichotomy between the core metallicity in CC and NCC clusters is already well known and consistent with previous works \citep[e.g.,][]{degrandi01,leccardi08b}.

The only significant metallicity evolution is measured in the cores of CC clusters, for which we measure $dZ/dz = -0.21 \pm 0.11$ Z$_{\odot}$. While only significant at the 2$\sigma$ level, it is worth noting that this evolution corresponds to, on average, $\sim$40\% of the metals in cool cores arriving between $z=1$ and $z=0$. This is significantly shallower than the 77$^{+7}_{-10}$\% evolution measured by \cite{ettori15} -- a point we will return to in the discussion.

This analysis has provided a clear picture of the metallicity evolution in galaxy clusters at $0.2 \lesssim z \lesssim 1.5$. We find no statistically significant evolution in outskirts of clusters, independent of the dynamical state of the core, and only marginal evidence for metallicity evolution in the inner regions of cool core clusters. We confirm substantial differences between CC and NCC cores in terms of the average metallicity, with CC clusters having significantly higher average metallicity. Outside of the core, CC and NCC clusters are indistinguishable based on their metallicity. Most importantly, we find that $>$60\% of the metals in the ICM were already in place at $z=1$ (with $>$95\% confidence). In the following section, we will place these results in the context of previous works and discuss their implications with regards to various enrichment scenarios.

\begin{deluxetable}{l c c c c}[h!]
\tablecaption{Average Metallicity and Metallicity Evolution}
\tablehead{
\colhead{Region} &
\colhead{Sel.} &
\colhead{$\left<Z[z=0.6]\right>$} &
\colhead{$dZ/dz$} &
\colhead{$\sigma_Z$}
\\
\colhead{} &
\colhead{} &
\colhead{[Z$_{\odot}$]} &
\colhead{[Z$_{\odot}$]} &
\colhead{[Z$_{\odot}$]} 
}
\startdata
\\
\multicolumn{5}{c}{\emph{Chandra Only}}\\

0.0 -- 0.15 R$_{500}$ &     & $0.32 \pm 0.03$ & $-0.04 \pm 0.10$ & $0.12 \pm 0.02$ \\
0.0 -- 0.15 R$_{500}$ &  CC & $0.39 \pm 0.03$ & $-0.21 \pm 0.11$ & $0.08 \pm 0.04$ \\
0.0 -- 0.15 R$_{500}$ & NCC & $0.24 \pm 0.05$ & $+0.03 \pm 0.18$ & $0.09 \pm 0.04$ \\
\\
0.15 -- 1.0 R$_{500}$ &     & $0.17 \pm 0.02$ & $-0.03 \pm 0.06$ & $0.06 \pm 0.01$ \\
0.15 -- 1.0 R$_{500}$ &  CC & $0.16 \pm 0.03$ & $+0.02 \pm 0.10$ & $0.06 \pm 0.03$ \\
0.15 -- 1.0 R$_{500}$ & NCC & $0.18 \pm 0.02$ & $-0.05 \pm 0.09$ & $0.06 \pm 0.02$ \\
\\
0.0 -- 1.0 R$_{500}$ &     & $0.22 \pm 0.02$ & $-0.06 \pm 0.06$ & $0.09 \pm 0.01$ \\
0.0 -- 1.0 R$_{500}$ &  CC & $0.25 \pm 0.03$ & $-0.06 \pm 0.10$ & $0.10 \pm 0.02$ \\
0.0 -- 1.0 R$_{500}$ & NCC & $0.19 \pm 0.02$ & $-0.09 \pm 0.08$ & $0.05 \pm 0.02$ \\

\\
\multicolumn{5}{c}{\emph{Chandra + XMM-Newton + Suzaku}}\\
0.0 -- 1.0 R$_{500}$ &     & $0.23 \pm 0.01$ & $-0.06 \pm 0.04$ & $0.08 \pm 0.01$ \\
\enddata
\tablecomments{Fitting parameters (normalization, slope, and scatter) from our maximum likelihood analysis described in \S4. The sample is divided up into subsamples by radial range (core, core-excised, and total) and by the cooling state of the core (cool core, non-cool core, combined). All uncertainties quoted are 1$\sigma$. The only subsample with non-negligible metallicity evolution is the cores of cool-core clusters, and this evolution is only significant at the $\sim$1$\sigma$ level.}
\label{table:mcmc}
\end{deluxetable}

\section{Discussion}
\subsection{Comparison to Previous Works}

There are a number of previous studies which have attempted to constrain the metallicity evolution of the ICM, both directly and indirectly \citep[e.g.,][]{balestra07,maughan08,anderson09,baldi12b,werner13,ettori15}. Conveniently, \cite{baldi12b} have compiled data from several of these earlier works, and expressed the metallicity evolution in a consistent way. To allow a direct comparison to these works, we will adopt the formalism of \cite{baldi12b} for the remainder of this section. These authors describe the evolution of the ICM metallicity as $Z \propto (1+z)^{-\gamma}$, with a normalization constant at $z=0.6$, leading to the expression:

\begin{equation} 
Z(z) = \frac{Z_{z=0.6}}{1.6^{-\gamma}}(1+z)^{-\gamma}~~~.
\end{equation}

\begin{figure}[ht]
\centering
\includegraphics[width=0.49\textwidth]{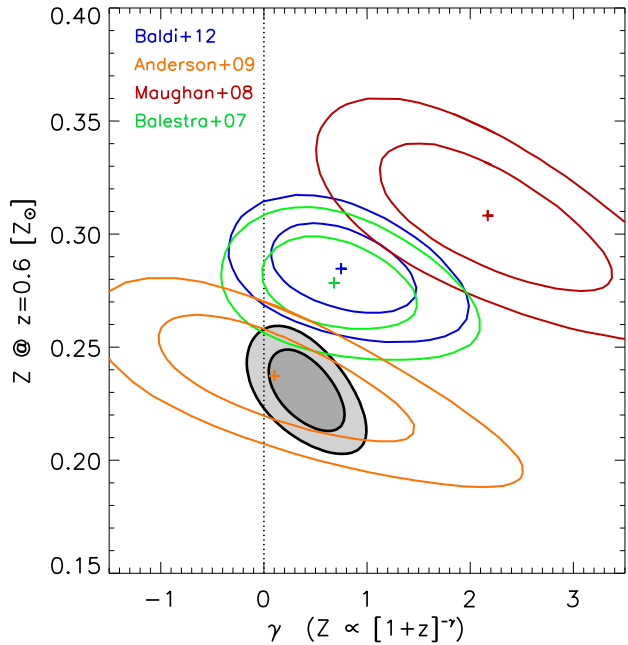}
\caption{Constraints on integrated ($r<R_{500}$) ICM metallicity evolution from this work compared to previous works. Colored contours, which enclose 68\% and 95\% confidence regions, are taken from \cite{baldi12b}. 
The grey contours show the significantly (factor of 2) improved constraints on the evolutionary parameter $\gamma$ from this work.
The lack of significant improvement in the normalization term stems from the fact that we are not actually measuring the metallicity more precisely than previous studies, nor do we have more clusters in our sample -- it is the redshift distribution of these clusters that leads to an improvement in the slope.}
\label{fig:mcmc_baldi}
\end{figure}

\noindent{}We re-run our analysis using the same model, for consistency. We also consider a new aperture, $0.15R_{500} < r < 0.5R_{500}$, which is similar to those used in previous works \citep[e.g.,][]{baldi12b,ettori15}.
In Figure \ref{fig:mcmc_baldi} we compare the results of this re-analysis to previous works \citep{balestra07,maughan08,anderson09,baldi12b,ettori15}. 
For clarity, we provide in Table \ref{table:baldi} two easily-interpreted quantities for each fit: the average metallicity predicted at $z=0$ and the predicted change in metallicity from $z=1$ to $z=0$. We derive these two estimates, and their uncertainties, based on the model fits provided in the literature, and compare to this work.
In general, the work presented here agrees well with results from the literature, with this work providing the tightest constraints to date on the metallicity evolution. Much of the discrepancy between works can be attributed to different apertures (see e.g., column 2 of Table \ref{table:baldi}) and different selection methods/criteria. 

\begin{deluxetable}{c c c c}[htb]
\tablecaption{Comparison to Literature}
\tablehead{
\colhead{Pub} &
\colhead{Extraction} &
\colhead{$\left<Z[z=0]\right>$} &
\colhead{$\Delta Z \big|_{z=1}^{z=0}$}
\\
\colhead{} &
\colhead{Radius} &
\colhead{[Z$_{\odot}$]} &
\colhead{[Z$_{\odot}$]}
}

\startdata
\multicolumn{4}{c}{\emph{Core-Included}} \\ \\
{\bf This work} & $0 < r < R_{500}$ & $0.28 \pm 0.04$ & $0.07 \pm 0.05$\\
B12 & $0 < r < 0.6R_{500}$ & $0.41 \pm 0.09$ & $0.17 \pm 0.12$\\
A09 & $0 < r < R_{S/N}$ & $0.27 \pm 0.11$ & $0.03 \pm 0.15$\\
M08 & $0 < r < R_{500}$ & $0.92 \pm 0.33$ & $0.72 \pm 0.36$\\
\\
\\
\multicolumn{4}{c}{\emph{Core-Excised}} \\ \\
{\bf This work} & $0.15R_{500} < r < R_{500}$ & $0.21 \pm 0.07$ & $0.05 \pm 0.09$\\
{\bf This work} & $0.15R_{500} < r < 0.5R_{500}$ & $0.23 \pm 0.06$ & $0.02 \pm 0.08$\\
E15 & $0.15R_{500} < r < 0.4R_{500}$ & $0.40 \pm 0.19$ & $0.16 \pm 0.24$\\
E15$^{*}$ & $0.15R_{500} < r < 0.4R_{500}$ & $0.30 \pm 0.20$ & $0.09 \pm 0.26$\\
B12 & $0.15R_{500} < r < 0.6R_{500}$ & $0.34 \pm 0.12$ & $0.12 \pm 0.15$\\
BLS07 & $0.15R_{vir} < r < 0.3R_{vir}$ & $0.40 \pm 0.10$ & $0.15 \pm 0.12$
\enddata
\tablecomments{Comparison of our evolutionary constraints on the ICM metallicity to those from the literature, assuming a fitting function of the form $Z \propto (1+z)^{-\gamma}$. All literature values here are derived from Table 4 of \cite{baldi12b}, with the exception of those from \cite{ettori15}. References are abbreviated to BLS07 \citep{balestra07}, M08 \citep{maughan08}, A09 \citep{anderson09}, B12 \citep{baldi12b}, and E15 \citep{ettori15}. The outer radius $R_{S/N}$ is based on a signal-to-noise criterion described in detail in \cite{anderson09}. We separate the comparison into literature measurements where the cluster core was excised, and those where it was not. In general, there is good agreement between this work and those of previous authors.
\\$^*$: Non-cool cores only
}
\label{table:baldi}
\end{deluxetable}

Considering the integrated metallicity (core included), the constraints on the evolutionary term, $\gamma$, are substantially improved by this work, compared to previous works. We find $\gamma = 0.41 \pm 0.25$, consistent with no evolution. For comparison, \cite{baldi12b} find $\gamma = 0.75 \pm 0.47$, which (for a fixed normalization) is consistent with anywhere from a 0--80\% change in the emission-weighted metallicity since $z=1$ (at 95\% confidence), compared to 0--40\% found in this study. There is excellent agreement between our work and those of \cite{anderson09} and \cite{baldi12b}, and some tension with the earlier results of \cite{maughan08}.

The tightest constraints prior to this work came from \cite{ettori15}, who utilized a sample of 83 galaxy clusters at $0.08 < z < 1.4$, all observed with \emph{XMM-Newton}. These clusters were drawn from the literature, so the selection effects are challenging to quantify precisely. For their full sample, combining metallicities measured at all radii, \cite{ettori15} found a metallicity normalization ($Z_0 = 0.70 \pm 0.12$ Z$_{\odot}$) that is significantly higher ($>$3$\sigma$) than our measurement. We do not show a direct comparison to this measurement in Figure \ref{fig:mcmc_baldi}, since it is not an integrated measurement but rather a two-dimensional fit to both radius and redshift. The high average metallicity found by this work for the full sample suggests that it may have a higher fraction of cool cores than is representative of the true population. The strong evolution ($\gamma=1.31\pm0.57$) measured by \cite{ettori15} may represent a redshift-dependent selection, with a higher-than-normal fraction of cool cores at low-$z$ and a lower-than-normal fraction high-$z$, consistent with other X-ray surveys. 

If, instead, single apertures are considered (e.g., $0.15R_{500} < r < 0.5$R$_{500}$), our analysis and that of \cite{ettori15} agree well, with measurements of $Z_0 = 0.23 \pm 0.06$ Z$_{\odot}$ and $Z_0 = 0.40 \pm 0.19$ Z$_{\odot}$, respectively. Further, if only non-cool cores are considered, this agreement is even better, with \cite{ettori15} finding $Z_0 = 0.30 \pm 0.20$ Z$_{\odot}$. The larger disagreement on the \emph{core-excised} metallicity when cool core clusters are included may be indicating an inability to fully excise the core for clusters at $z>1$ using data from \emph{XMM-Newton}. For a typical cluster in our sample, at $z=1.2$, 0.15R$_{500} \sim 12^{\prime\prime}$, which is similar in size to the on-axis FWHM of \emph{XMM-Newton}. Thus, we would expect the core-excised metallicity measurements of high-$z$ cool core clusters based on \emph{XMM-Newton} data to be biased slightly high.

In general, our results agree well with previous works and, in cases where there is disagreement, it is clear how both selection biases and differences in analyses could conspire to account for any discrepancies. It is worth repeating that the results presented here also agree well with the indirect measurement of non-evolving ICM metallicity from \cite{werner13} and \cite{simionescu15}. These works found azimuthally-uniform metallicities in the outskirts of Perseus and Virgo, with $\left<Z\right> = 0.21 \pm 0.01$ Z$_{\odot}$ and $0.16 \pm 0.03$ Z$_{\odot}$, respectively (assuming the same solar abundances). These azimuthally-uniform profiles were taken as evidence for a lack of metallicity evolution, since mixing times are long at large radii. These works compare favorably with ours, both in terms of the average metallicity in the outskirts of clusters at $z=0$ ($0.21 \pm 0.07$ Z$_{\odot}$ in this work), and in terms of the implied weak evolution of the ICM metallicity.

\subsection{Dependence of Results on Data Quality}

The majority of the data used in this program comes from a series of large programs aimed at obtaining $\sim$2000 X-ray counts per cluster. As such, the median number of counts in the 96 \emph{Chandra} spectra used in this work is $\sim$1800. There are, however, 7 clusters in this sample with deep ($>$10$^4$ total counts) X-ray observations. These 7 systems account for 82\% of the total signal from the \emph{Chandra} sample of 96 clusters. To test how sensitive our analysis is to the contributions of a few systems, we determine the best-fitting parameters ($\left<Z(z=0.6)\right>$, $dZ/dz$, $\sigma_Z$) for the full 96-cluster \emph{Chandra} sample, as well as a subsample excluding these 7 high-S/N systems. 

\begin{figure}[t]
\centering
\includegraphics[width=0.49\textwidth]{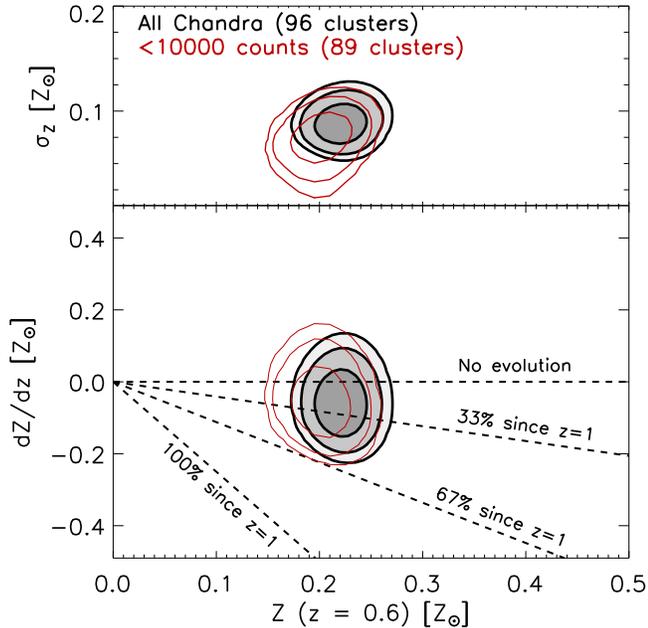}
\caption{Constraints on the evolution of the metallicity within $r<R_{500}$ for the full sample of 96 \emph{Chandra}-observed clusters (grey) and a subsample excluding the 7 highest signal-to-noise systems. The exclusion of these 7 systems, which comprise 82\% of the total signal, does not have a strong effect on the outcome of this analysis, shifting the normalization and scatter by $\sim$1$\sigma$, and leaving the slope relatively unchanged. To first order, the effect of including a small number of high signal-to-noise measurements is to improve constraints on the measured scatter.}
\label{fig:mcmc_s2n}
\end{figure}

In Figure \ref{fig:mcmc_s2n} we show constraints on the best-fitting metallicity evolution model for the full \emph{Chandra} sample in an aperture of $r<R_{500}$. Excluding the 7 highest S/N systems, which comprise 82\% of the total X-ray signal, from this analysis has only a marginal effect. We find shifts of $1.1\sigma$, $0.04\sigma$, and $1.0\sigma$ in the normalization, slope, and scatter of the metallicity evolution. The highest signal-to-noise systems improve the constraints on the scatter by 27\%, while providing relatively little improvement to the normalization and slope. This test confirms that the primary effect of including a small number of precise measurements in this analysis is to improve constraints on the scatter, and that individual systems are not driving the measured evolution.

\subsection{The Origin of ICM Metal Enrichment}

The results presented here strongly suggest that there has been relatively little change ($<$40\%) in the ICM metallicity since $z=1$, consistent with observations of the outskirts of galaxy clusters, which show remarkable uniformity with azimuth \citep{werner13, simionescu15}. These metals were likely formed in a mix of core-collapse supernovae (CCSN) and type Ia supernovae -- the fact that the relative metal abundances are also uniform with azimuth imply that enrichment from both of these sources happened relatively early.

The relative metal abundances in the outskirts of clusters indicate that the enrichment was dominated by CCSN \citep[see review by][]{werner08}, with only a small fraction of the metals coming from type Ia SNe \citep[12--37\%;][]{simionescu15}. The data presented here are consistent with a picture in which the bulk of the metals present in the ICM today were produced from CCSN during an early stage of rapid star formation, and removed from their host galaxies via starburst-driven winds \citep[see review by][]{veilleux05}. In the centers of cool core clusters, where we observe a peak in the metallicity, it is thought that the metals are produced via type Ia SNe in the central BCG \citep{degrandi04}. 

In the centers of cool core clusters, we estimate that emission-weighted metallicity has increased by $40\pm20$\% in the intervening time since $z=1$ (see Figure \ref{fig:mcmc_chandra}). Assuming our best estimate of the metal enrichment rate in the centers of cool core clusters clusters (Table \ref{table:mcmc}), coupled with no evolution in the outer (core excised) metallicity, it would take $\sim$10 Gyr to build the observed present-day central excess of metals in the centers of cool core clusters. Thus, if our maximal likelihood estimate of the core and outer metallicity evolution is correct, we would expect nearly flat metallicity profiles in galaxy clusters at $z\gtrsim1.5$.
%


One possible avenue for the outskirts of galaxy clusters to become metal-enriched is for powerful radio jets from the central cluster galaxy to push metals out from the high-metallicity core to the low-metallicity outskirts. Such metal-enriching outflows have been predicted in simulations \citep[e.g.,][]{gaspari11} and observed in multiple systems \citep[e.g.,][]{kirkpatrick11}. In the systems studied here, we find that the average metallicity outside of the core at $z=0.6$ is $0.16 \pm 0.03$ Z$_{\odot}$ and $0.18 \pm 0.02$ Z$_{\odot}$ for cool core and non-cool core clusters respectively (see Table \ref{table:mcmc}). If we assume that radio outflows are restricted to cool core clusters \cite[e.g.,][]{sun09b}, the similarity in the core-excised metallicity suggests that only a small fraction of the total metals can be transported outside of the core, on average. Given the parameters in Table \ref{table:mcmc}, we can say (with 95\% confidence) that $<$10\% of the metallicity in an average cool core cluster is transferred outside of the core by AGN feedback -- any higher fraction would be observable given our uncertainties. This is consistent with the results presented by \cite{kirkpatrick11}, who showed that Fe-rich outflows extended to $\sim$0.15R$_{500}$, but not beyond, and with simulations showing that AGN feedback can not break self-similarity outside of cluster cores \citep{gaspari14}.

\section{Summary}

We present an analysis of the ICM metallicity in 153 mass-selected galaxy clusters spanning $0 < z < 1.5$. This sample of clusters, observed with the \emph{Chandra}, \emph{XMM-Newton}, and \emph{Suzaku} X-ray satellites, provides the best constraints to date on the mean, scatter, and evolution of the ICM metallicity. The main results of this work are summarized as follows:

\begin{itemize}

\item We find no evidence for evolution in the \emph{global} ($r < R_{500}$) ICM metallicity of clusters spanning $0 < z < 1.5$. We report that the emission-weighted metallicity has not changed by more than 40\% since $z=1$ (at 95\% confidence).

\item We find a $>$3$\sigma$ difference between the core ($r<0.15R_{500}$) metallicities of cool core and non-cool core clusters, consistent with earlier works. 

\item We find only a weak ($\sim$33\%) radial increase in metallicity toward the centers of non-cool core clusters, compared to a $>$140\% increase in the centers of cool core clusters. 

\item We find no evidence for metallicity evolution in the cores of non-cool core clusters ($dZ/dz = -0.03 \pm 0.06$).

\item Our best estimate ($dZ/dz = -0.21 \pm 0.11$) suggests that the metallicity enhancement observed in the centers of low-$z$ cool cores may have been built slowly over the past $\sim$10 Gyr.

\item We find no evidence for evolution in the core-excised ($0.15R_{500} < r < R_{500}$) ICM metallicity, and no difference between the core-excised metallicity between cool core and non-cool core clusters. This implies that radio jets originating in the central cluster galaxy can not move a significant ($>$10\%) fraction of the metals beyond $0.15R_{500}$ in cool core clusters.

\end{itemize}

These data imply that the bulk of the metals in the ICM were incorporated at early times ($z\gtrsim1.2$), most likely during the peak of star formation at $z\sim2$. This work represents a total commitment of 9.1 Ms from three of the most sensitive X-ray telescopes in orbit. 
As such, a significant improvement on this work is unlikely to come from deeper observations with current-generation telescopes, with the exception of potentially confirming a lack of highly-enriched cool cores at $z>1$. A more precise accounting of the enrichment history of the ICM awaits next-generation observatories, such as \emph{Athena} and \emph{X-ray Surveyor}, combined with samples of clusters at $z\sim2$ which should be available with the next generation of SZ experiments \citep[e.g., SPT-3G;][]{benson14}.


\section*{Acknowledgements} 
Much of this work was enabled by generous GTO contributions from Steve Murray, and was in progress at the time of his untimely death in 2015. He was a valued member of the Center for Astrophysics and a strong supporter of SPT science - he will be greatly missed by all of us.
M.\ M.\ acknowledges support by NASA through contracts GO4-15122A and GO5-16141X (Chandra), and support by NASA through a Hubble Fellowship grant HST-HF51308.01-A awarded by the Space Telescope Science Institute, which is operated by the Association of Universities for Research in Astronomy, Inc., for NASA, under contract NAS 5-26555. 
E.\ B.\ acknowledges support by NASA through contract NNX123AE77G.
The South Pole Telescope program is supported by the National Science Foundation through grants ANT-0638937 and PLR-1248097. 
This work was partially completed at Fermilab, operated by Fermi Research Alliance, LLC under Contract No. De-AC02-07CH11359 with the United States Department of Energy.
Partial support is also provided by the NSF Physics Frontier Center grant PHY-0114422 to the Kavli Institute of Cosmological Physics at the University of Chicago, the Kavli Foundation, and the Gordon and Betty Moore Foundation. 
Support for X-ray analysis was provided by NASA through Chandra Award Numbers 12800071, 12800088, and 13800883 issued by the Chandra X-ray Observatory Center, which is operated by the Smithsonian Astrophysical Observatory for and on behalf of NASA. 
Argonne National Laboratory's work was supported under the U.S. Department of Energy contract DE-AC02-06CH11357
C.\ R.\ acknowledges support from the Australian Research Council?s Discovery Projects scheme (DP150103208).
%



\end{document}